# Yin-Yang vortex on $UTe_2$ (011) surface

Ruotong Yin[1,#], Yuanji Li[1,#], Zengyi Du[2,#], Dengpeng Yuan[3,#], Shiyuan Wang[1], Jiashuo Gong[1], Mingzhe Li[1], Ziyuan Chen[1], Jiakang Zhang[1], Yuguang Wang[1], Ziwei Xue[3], Xinchun Lai[3], Shiyong Tan[3,*], Da Wang[4,5,*], Qiang-Hua Wang[4,5,*], Dong-Lai Feng[1,6,*], Ya-Jun Yan[1,2,*]

[1]*Hefei National Research Center for Physical Sciences at the Microscale and Department of Physics, University of Science and Technology of China, Hefei, 230026, China*
[2]*Hefei National Laboratory, University of Science and Technology of China, Hefei, 230026, China*
[3]*Science and Technology on Surface Physics and Chemistry Laboratory, Mianyang, 621907, China*
[4]*National Laboratory of Solid State Microstructures and School of Physics, Nanjing University, Nanjing, 210093, China*
[5]*Collaborative Innovation Center of Advanced Microstructures, Nanjing University, Nanjing, 210093, China*
[6]*National Synchrotron Radiation Laboratory, School of Nuclear Science and Technology, and New Cornerstone Science Laboratory, University of Science and Technology of China, Hefei, 230026, China*

[#] ***Those authors contributed equally to this work.***
*E-mails: sytan4444@163.com, dawang@nju.edu.cn, qhwang@nju.edu.cn, dlfeng@ustc.edu.cn，yanyj87@ustc.edu.cn;

**$UTe_2$ is a promising candidate for spin-triplet superconductor, yet its exact superconducting order parameter remains highly debated. Here, via scanning tunneling microscopy/spectroscopy, we observe a novel type of magnetic vortex with distinct dark-bright contrast in local density of states on $UTe_2$ (011) surface under a perpendicular magnetic field, resembling the conjugate structure of Yin-Yang diagram in Taoism. Each Yin-Yang vortex contains a quantized magnetic flux, and the boundary between the Yin and Yang parts aligns with the crystallographic *a*-axis of $UTe_2$. The vortex states exhibit intriguing behaviors — a sharp zero-energy conductance peak exists at the Yang part, while a superconducting gap with pronounced coherence peaks exists at the Yin part, which is even sharper than those measured far from the vortex core or in the absence of magnetic field. By theoretical modeling, we show that the Yin-Yang vortices on $UTe_2$ (011) surface can be explained by the asymmetric vortex-derived local distortion of the zero-energy surface states associated with spin-triplet pairing with appropriate *d*-vectors. Therefore, the observation of Yin-Yang vortex confirms the spin-triplet pairing in $UTe_2$ and imposes constraints on the candidate *d*-vector for the spin-triplet pairing.**

## I. INTRODUCTION

Since the discovery of superfluidity in $^3$He five decades ago [1], physicists have been seeking more spin-triplet superconductors both to understand this exotic state and to harness their inherent topologically protected surface states for quantum computing and energy-efficient electronics [2]. $UBe_{13}$ and $UPt_3$ are possible candidates with many unconventional superconducting properties [3,4], but their low superconducting critical temperature ($T_c$) limits extensive experimental studies. Since 2019, $UTe_2$ has emerged as a new promising candidate for spin-triplet superconductors, the relatively high $T_c$ and recently improved sample quality have enabled the acquisition of further evidence supporting its triplet pairing nature [5-27], including the large and highly anisotropic upper critical field that exceeds the Pauli limit [5,6], weak temperature dependence of the Knight shift across $T_c$ (Refs. [8,18,23,26]), multiple reentrant superconducting phases with increasing magnetic field [7,9,11,15], chiral in-gap states at step edges of $UTe_2$ (011) surface [14], etc. However, the precise form of the superconducting order parameter in $UTe_2$ remains highly debated experimentally [14,18,20,23,26,28,29], and the existence of the associated topologically protected surface states has yet to be definitively confirmed.

$UTe_2$ crystallizes in a body-centered orthorhombic structure with $D_{2h}$ point group symmetry. In the presence of strong spin-orbit coupling, four distinct single-component spin-triplet superconducting order parameters are allowed: $A_u$ with $\boldsymbol{d}_k = (\alpha k_a, \beta k_b, \gamma k_c)$, $B_{1u}$ with $\boldsymbol{d}_k = (\beta k_b, \alpha k_a, 0)$, $B_{2u}$ with $\boldsymbol{d}_k = (\gamma k_c, 0, \alpha k_a)$, and $B_{3u}$ with $\boldsymbol{d}_k = (0, \gamma k_c, \beta k_b)$ (Ref. [19]), here $(\alpha, \beta, \gamma)$ are real space components of the $\boldsymbol{d}$-vector (defined in spin space). Among them, $A_u$ generally corresponds to a fully gapped superconducting state, and gap nodes can only exist when there is an accidental zero in the pairing parameters $(\alpha, \beta, \gamma)$; In contrast, gap nodes exist generally for $B_{1u}$, $B_{2u}$, and $B_{3u}$ states along the crystalline ***c***-, ***b***-, and ***a***-axes, respectively. In principle, multicomponent superconductivity may arise from linear combinations of these four order parameters, breaking time reversal symmetry (TRS) [30]. However, the latest results of Kerr and muon spin relaxation ($\mu$SR) measurements on high-quality $UTe_2$ with $T_c \sim 2$ K rule out this possibility [31,32], in contrast to early evidence of TRS breaking in samples with lower $T_c \sim 1.6$ K (Refs. [20,21]). Thus, the superconducting state in $UTe_2$ is most likely time-reversal invariant and possesses a single component. Furthermore, thermal conductivity [12,33], specific heat [16,24] and penetration depth [28] measurements demonstrate the presence of nodal quasiparticles in its superconducting state, favoring $B_{1u}$, $B_{2u}$, and $B_{3u}$ over fully gapped $A_u$ state. More order-parameter sensitive experiments are needed to distinguish these pairing symmetries.

The behaviors of magnetic vortex can provide critical clues on the nature of superconducting order parameters. For instance, in an anisotropic or nodal superconductor, vortex structures may exhibit complex spatial patterns or nodal features [34,35]; In topological superconductors, Majorana zero modes are expected to emerge at the vortex core [36-38]. However, due to the previously limited sample quality, direct imaging of magnetic vortex and study of vortex states in $UTe_2$ remains elusive. Here, we use scanning tunneling microscopy/spectroscopy (STM/STS) with ultralow-temperature and high-magnetic-field capabilities to investigate the magnetic vortex properties on the (011) surface of high-quality $UTe_2$ crystals. We directly observe a novel type of vortex featuring a Yin-Yang-like spatial contrast, and combined with theoretical analysis, we attribute its appearance to the interplay between the zero-energy surface states associated with spin-triplet superconductivity and the vortex, the latter of which generates local distortion of the surface states asymmetrically with respect to the vortex core.

## II. RESULTS

### A. Basic topographic and spectral properties of $UTe_2$ (011) surface

The crystal structure of $UTe_2$ is shown in Fig. 1(a), where the commonly reported easy-cleave plane, the (011) plane, is indicated by the yellow plane. The (011) plane consists of chains of Te1, Te2 and U atoms aligned along crystallographic ***a***-axis, with their atomic heights arranged in descending order (Fig. 1(b)). $UTe_2$ crystals used in this study are synthesized via the molten salt flux method, which show a $T_{c,0}$ of ~ 2 K and residual resistance ratio of ~ 82 (Supplementary note 1 of Supplemental Material (SM) [39]). Figure 1(c) shows the typical topographic image of the exposed surface of $UTe_2$ after cleavage, exhibiting chain-like structures along ***a***-axis; A higher-resolution image (Fig. 1(d)) resolves two alternating atomic chains with different heights, which are assigned as Te1 and Te2, respectively. The measured interatomic spacing along Te1 chain and the interchain spacing between adjacent Te1 chains are $a$ = 4.05 Å and $b^*$ = 7.6 Å, consistent with the lattice parameters of $UTe_2$ (011) plane (Refs. [5,45]), as well as the previously reported STM results on $UTe_2$ (011) surface[14,27,46-49]. Here we define the perpendicular direction to ***a***-axis on $UTe_2$ (011) surface as $\boldsymbol{b}^*$-axis to distinguish it from crystallographic ***b***-axis.

Figures 1(e) and 1(f) show the spatial evolution of superconducting gap spectra along Te2 chain (cut #1) and perpendicular to it (cut #2). The superconducting gap is obvious and homogeneous along Te2 chain, with a gap size of ~ 0.25 meV judging from the coherence peak positions and an averaged gap depth of ~ 15% defined by 1 - $g$(0 meV)/$g$(1 meV), with $g(E)$ representing the d$I$/d$V$ value at energy $E$. By contrast, the superconducting gap varies significantly along $\boldsymbol{b}^*$-axis, consistent with previous reports[14]. It is noteworthy that the observed shallow superconducting gap reflects the intrinsic property of $UTe_2$, as the superconducting gap of Al polycrystal with $T_c$ ~ 1.2 K measured under the same conditions is sharp, showing 100% gap depth (Supplementary note 2 of SM [39]). Taking account of existing studies[12,16,20,23,24,28,29,33], this might be explained by considering that $UTe_2$ is a spin-triplet superconductor with intrinsic gap nodes and gapless surface states, which contribute to the residual zero-energy density of states (DOS).

### B. Yin-Yang vortex on $UTe_2$ (011) surface

Subsequently, we measure the magnetic vortex properties on $UTe_2$ (011) surface. Figure 2(a) shows the topographic image of a selected sample region of 190 × 190 $nm^2$, and the raw vortex maps, $g(\mathbf{r}, E)$ = d$I$/d$V(\mathbf{r}, E)$, collected in this field of view under different perpendicular magnetic fields ($\mathbf{B}_\perp$) are displayed in Fig. S3 of SM [39], where unique vortex patterns can be resolved. To enhance contrast, we defined normalized vortex maps $C(\mathbf{r}$, 0.25 meV) = $g(\mathbf{r}$, 0 meV)/$g(\mathbf{r}$, 0.25 meV), that is dividing the raw vortex map at zero-energy by that at coherence peak energy, which are presented in Figs. 2(b)-2(g). Without magnetic field, the local DOS (LDOS) distribution exhibits some spatial inhomogeneity but no discernible patterns can be resolved (Fig. 2(b)). Under $\mathbf{B}_\perp$= 0.25 T, bright stripes appear along ***a***-axis as indicated by the blue arrow in Fig. 2(c), accompanied by a peculiar dark core with suppressed LDOS on its right side (white arrow). Such vortex pattern with paired bright-dark cores is intrinsic to $UTe_2$ (011) surface, which remains unchanged when the STM scanning direction is changed or the direction of $\mathbf{B}_\perp$ is reversed, as discussed in Supplementary note 4 of SM [39]. Moreover, it persists to higher $\mathbf{B}_\perp$ values (Figs. 2(d)-2(g)), with its number increasing proportionally with increasing $\mathbf{B}_\perp$, and finally becomes invisible at $\mathbf{B}_\perp$ > 8 T (Fig. S3 of SM [39]). Figure 2(h) summarizes the number of paired bright-dark cores as a function of $\mathbf{B}_\perp$, which fits well with the theoretical value of $n = \frac{\mathbf{B}_\perp \cdot \mathbf{S}}{\Phi_0}$, with $S$ and $\Phi_0$ are the scan area and the magnetic flux quantum, demonstrating that each paired bright-dark cores contains a quantized magnetic flux. To our knowledge, there is no prior report of such kind of magnetic vortex, which we term "Yin-Yang" vortex in this study, as its dark-bright parts resemble the conjugate structure of the Yin-Yang diagram.

Figure 3(a) shows the detailed distribution of a single Yin-Yang vortex measured at $\mathbf{B}_\perp$ = 0.5 T, the LDOS contrast between the Yang and Yin parts of the vortex is clearly resolved. Typical d$I$/d$V$ spectra measured at the Yang part (point A in Fig. 3(a) with higher zero-energy DOS), Yin part (point

B in Fig. 3(a) with lower zero-energy DOS) of the vortex, locations away from the vortex, and that without magnetic field are plotted together in Fig. 3(b) for comparison. Far away from the vortex core, a shallow gap with subtle coherence peaks is observed, resembling that observed in the absence of magnetic field, which is probably induced by the inherent nodal gap structure and gapless surface states of spin-triplet superconductivity in $UTe_2$ as discussed above. In the vortex core, the d$I$/d$V$ spectrum at the Yang part exhibits a sharp zero-energy conductance peak (ZECP) with a full width at half maximum (FWHM) of ~ 0.2 meV, persisting to tens of nanometers along $\boldsymbol{a}$-axis without obvious splitting (Supplementary note 5 of SM [39]), which may result from the superposition of Caroli-de-Gennes-Matricon (CdGM) bound states and gapless surface states; whereas in strong contrast, the d$I$/d$V$ spectrum at the Yin part shows a superconducting gap with pronounced coherence peaks that are even sharper than those measured far from the vortex core or under zero magnetic field, which is puzzling and suggests reduced contribution of CdGM states and gapless surface states near Fermi level.

Figures 3(c) and 3(d) display typical line profiles of the vortex acquired along cuts #3 and #4 in Fig. 3(a). In Fig. 3(c), a single-exponential function was used to fit the data and extract the Ginzburg-Landau coherence length along $\boldsymbol{a}$-axis ($\xi_a$), the obtained results for multiple vortices are shown in Fig. 3(e), yielding an averaged $\xi_a$ = 15.1 ± 4.8 nm. In Fig. 3(d), a double-exponential fitting model was applied to separately extract the coherence lengths of the vortex's Yang and Yin parts ($\xi_{\mathrm{Yang}}$ and $\xi_{\mathrm{Yin}}$) along $\boldsymbol{b}^*$-axis, as well as the center-to-center distance between them, $d_{\mathrm{Yin\text{-}Yang}}$. The corresponding results for multiple vortices, along with the values of $\xi_{\mathrm{Yang}} + \xi_{\mathrm{Yin}}$, are presented in Figs. 3(f) and 3(g), and the averaged values of $\xi_{\mathrm{Yang}}$, $\xi_{\mathrm{Yin}}$, $\xi_{\mathrm{Yang}} + \xi_{\mathrm{Yin}}$, and $d_{\mathrm{Yin\text{-}Yang}}$ are approximately 4.5 ± 1.4 nm, 7.5 ± 2.1 nm, 12.0 ± 2.7 nm, and 10.6 ± 2.7 nm, respectively (see Supplementary note 6 of SM in Ref. [39] for more fitting details). Two notable features emerge: (i) $\xi_{\mathrm{Yang}}$ is significantly smaller than $\xi_{\mathrm{Yin}}$, and intriguingly, (ii) $d_{\mathrm{Yin\text{-}Yang}}$ approximates the sum $\xi_{\mathrm{Yang}}$ + $\xi_{\mathrm{Yin}}$. Taking $\xi_{\mathrm{Yang}} + \xi_{\mathrm{Yin}}$ as the coherence length along $\boldsymbol{b}^*$-axis, $\xi_{\mathrm{b}^*}$, the resulting ratio of $\xi_a$:$\xi_{\mathrm{b}^*}$ is ~ 1.3, which is likely due to the anisotropy of Fermi surface and superconducting gap structure.

### C. Theoretical analysis

Our experiments reveal a novel type of magnetic vortex pattern and vortex core states that deviates from the conventional vortex paradigm. To investigate whether this unique vortex pattern arises from the spin-triplet pairing nature of $UTe_2$, we performed numerical simulations for the vortex structures on $UTe_2$ (011) surface under the aforementioned four types of spin-triplet pairing symmetries, $A_u$, $B_{1u}$, $B_{2u}$ and $B_{3u}$, by considering a *normalized* $\boldsymbol{d}$-vector with $\alpha^2 + \beta^2 + \gamma^2 = 1$. To mimic the nearly square-shaped cylindrical Fermi surfaces of $UTe_2$ proposed in previous theoretical and experimental studies[10,13,25,50-53], we construct a single-orbital tight-binding model on the cubic lattice, to obtain a cylindrical squarish Fermi surface as shown in Fig. 4(a). Then we solve the Bogoliubov-de Gennes (BdG) Hamiltonian in the presence of a single vortex for the four types of spin-triplet pairing symmetries on the tilted lattice with (011) surface, with the technical details presented in Supplementary notes 7A-7C of SM [39]. In Supplementary note 7D of SM [39], a series of zero-energy LDOS on the (011) surface for different values of $(\alpha, \beta, \gamma)$ are present, as well as their spatial LDOS profiles along the $\boldsymbol{b}$*-axis cut passing through the vortex center. By comparing these results with our experiments, we find that the Yin-Yang pattern can occur for all the four types of spin-triplet pairings when the pairing parameters $(\alpha, \beta, \gamma)$ lie within the yellow shaded strip and its thick gray boundaries shown in Fig. 4(b). Four typical results of the zero-energy LDOS exhibiting the Yin-Yang pattern are plotted in Figs. 4(c)-4(f). As a comparison, we also considered the case of *s*-wave pairing (Fig. S11 of SM [39]), which does not exhibit the Yin-Yang pattern. Therefore, we attribute the Yin-Yang pattern to the spin-triplet pairing in $UTe_2$.

In order to further understand the underlying physical mechanism for this new phenomenon, we consider an extreme case with exactly square-shaped cylindrical Fermi surface such that the Fermi velocities exist only along ***a***- and ***b***-axes. We adopt the quasi-classical picture (Supplementary note 7E of SM [39]), which is valid when the Fermi wavelength is much shorter than the superconducting coherence length $\xi$. The LDOS $\rho(\omega, \mathbf{r})$ (proportional to the $dI/dV$ value in STM) is contributed by summing the classical paths along different (only two) Fermi velocities $\boldsymbol{v}_F$ on the Fermi surface, i.e. $\rho(\omega, \mathbf{r}) = \oint_{\boldsymbol{k}_F} \rho_{\boldsymbol{k}_F}(\omega, \mathbf{r})$. We consider five typical paths L1 to L5 along ***b***-axis as shown in Fig. 4(g). As far as the pairing potential (a combination of the macroscopic order parameter and the microscopic ***d***-vector) has a sign change on the incoming and outgoing directions (relative to the boundary) on the same path, there is at least one zero-energy edge mode [44] (see Supplementary note 7F of SM [39] for details), as depicted in Fig. 4(h) for mapping the path to a full chain to get a better understanding. For path L1, a zero-energy edge mode exists at the endpoint P1 (red curve). For path L5, there are two zero-energy modes, one at the endpoint P5 as the edge mode (red curve, the same as L1), and the other near the vortex center as the vortex bound state (orange curve). From L2 to L4, due to the drop and/or sign-reversal of the pairing potential, the wave function of the zero-energy mode (magenta curves) is distorted and more extensive in the evanescent direction. By normalization, the amplitude of the wave function at the boundary (contributing to zero-energy LDOS) drops from P2 to P4, resulting the suppressed surface LDOS asymmetrically with respect to the vortex core (e.g., P4 is suppressed deeper than P2). On the other hand, there are no contributions to the surface LDOS from the edge modes on the paths along ***a***-axis, since the ends are far from the vortex. Such paths do contribute bound states near the vortex, but they contribute to the LDOS symmetrically. Combining the above two factors, we gain a qualitative understanding of the Yin-Yang pattern of vortex bound states on $UTe_2$ (011) surface.

## III. DISCUSSION AND CONCLUSIONS

Based on the above quasi-classical picture, we gain a qualitative understanding that the Yin-Yang vortex pattern is caused by the interplay between the zero-energy surface states and the vortex in the case of cylindrical Fermi surface of $UTe_2$. Therefore, it is strong evidence for the existence of zero-energy surface states associated with the spin-triplet superconductivity on $UTe_2$ (011) surface, which requires a nonzero $k_b$ or $k_c$ component in the ***d***-vector (see Supplementary note 7B of SM [39] for details). This condition itself cannot rule out any one of the four pairing candidates, but according to our simulation results shown in Fig. 4(b), we find that the yellow shaded strip where Yin-Yang vortex pattern emerges is far away from the intersection points of the sphere and the coordinate axes, which does impose a strong constraint on the pairing parameters — at least two of $(\alpha, \beta, \gamma)$ must be nonzero and of comparable magnitudes. Given that the presence of nodal quasiparticles in the superconducting state of $UTe_2$ is extensively demonstrated [12,33], an additional constraint follows — at least one of $(\alpha, \beta, \gamma)$ must be zero. Therefore, there must be *one and only one zero* in the coefficients $(\alpha, \beta, \gamma)$ in the spin-triplet pairing functions, i.e. only on the thick gray boundaries of the yellow shaded strip in Fig. 4(b), which belong to $B_{1u}$, $B_{2u}$ and $B_{3u}$ with nodal gaps or to $A_u$ with accidental gap nodes. Further determination of the directions of the nodes and/or the ***d***-vector is needed to help pin down the pairing symmetry of $UTe_2$.

In conclusion, we report the observation of a novel type of magnetic vortex featuring a Yin-Yang-like spatial contrast on $UTe_2$ (011) surface, and combined with theoretical modeling, we attribute its appearance to the asymmetric vortex-derived local distortion of the zero-energy surface states associated with spin-triplet pairing with appropriate *d*-vectors. Therefore, the observation of Yin-Yang vortex confirms the spin-triplet pairing in $UTe_2$ and imposes constraints on the candidate *d*-vector for the spin-triplet pairing.

## IV. METHODS

### A. Sample synthesis and characterization

High-quality $UTe_2$ single crystals were synthesized via the molten salt flux (MSF) technique employing an equimolar NaCl-KCl mixture (99.99% purity, Alfa Aesar) as a flux. Uranium metal pieces (mass < 0.4 g) were initially etched in nitric acid to eliminate surface oxides. Under an argon atmosphere in a glovebox, tellurium pieces (Te, 99.999% purity, Alfa Aesar) were combined with uranium at a U:Te molar ratio of 1:1.65, along with NaCl-KCl flux at a U:salt molar ratio of 1:60. The reactants were loaded into a carbon crucible lined with quartz wool to prevent material loss during heating. Then the crucible was placed in a quartz tube and heated to 180 °C under high vacuum ($< 5\times10^{-4}$ Pa) for dehydration. Subsequently, the ampoule was vacuum-sealed and placed in a box furnace, where the mixture was heated to 950 °C over 24 hours and maintained at this temperature for 48 hours. It was then cooled gradually at a rate of 0.03 °C/min to 650 °C, held at this temperature for 48 hours, and finally cooled to ambient temperature naturally. After the growth process, the ampoules and crucibles were mechanically cleaved, and bulk $UTe_2$ crystals were manually collected and stored under an argon atmosphere to prevent oxidation.

Temperature dependent resistivity measurements were conducted in a Quantum Design DynaCool Physical Properties Measurement System (PPMS-9T), by using a standard four-probe configuration.

### B. STM measurements

$UTe_2$ crystals were mechanically cleaved at 80 K in ultrahigh vacuum with a base pressure better than $2 \times 10^{-10}$ mbar and immediately transferred into a UNISOKU-1600 STM. Pt-Ir tips were used for STM measurements after being treated on a clean Au (111) substrate. The $dI/dV$ spectra were collected by a standard lock-in technique with a modulation frequency of 973 Hz and a modulation amplitude $\Delta V$ of 50-100 μV. The data in the main text were collected at ~ 40 mK with an effective electron temperature $T_{\mathrm{eff}}$ of ~ 170 mK.

*Note added.* Recently, we became aware of another two works by Yang *et al.* [54] and Sharma *et al.* [55] reporting similar magnetic vortex structure and vortex states in $UTe_2$ (011) surface by STM measurements.

## ACKNOWLEDGMENTS

We thank Prof. Ziqiang Wang for helpful discussions. This work is supported by National Key R&D Program of the MOST of China (Grants No. 2023YFA1406304 (Y.J.Y.), No. 2022YFA1403201 (D.W.)), the National Natural Science Foundation of China (Grants No. 12374140 (Y.J.Y.), No. 12374147 (Q.H.W.), No. 12274205 (D.W.), No. 92365203 (Q.H.W.), No.

U23A20580 (S.Y.T.), No. 12494593 (Y.J.Y.)), the Innovation Program for Quantum Science and Technology (Grant No. 2021ZD0302803 (D.L.F.)), the New Cornerstone Science Foundation (D.L.F.), Sichuan Science and Technology Progr am (2025NSFJQ0040 (S.Y.T.)).

$UTe_2$ single crystals were grown by D. Y., Z. X., S. T. and X. L.; STM measurements were performed by R. Y. and Y. L.; Data analysis was performed by R. Y., Y. L., S. W., J. G., M. L., Z. C., J. Z., Y. W., Z. D. and Y. Y.; numerical simulation was done by D. W. and Q. W.; Z. D., D. W., Y. Y. and D. F. coordinated the whole work and wrote the manuscript. All authors have discussed the results and the interpretation.

## DATA AVAILABILITY

The main data supporting the findings of this study are available within the article and its Supplementary Information files. All the raw data generated in this study are available from the corresponding author upon request.

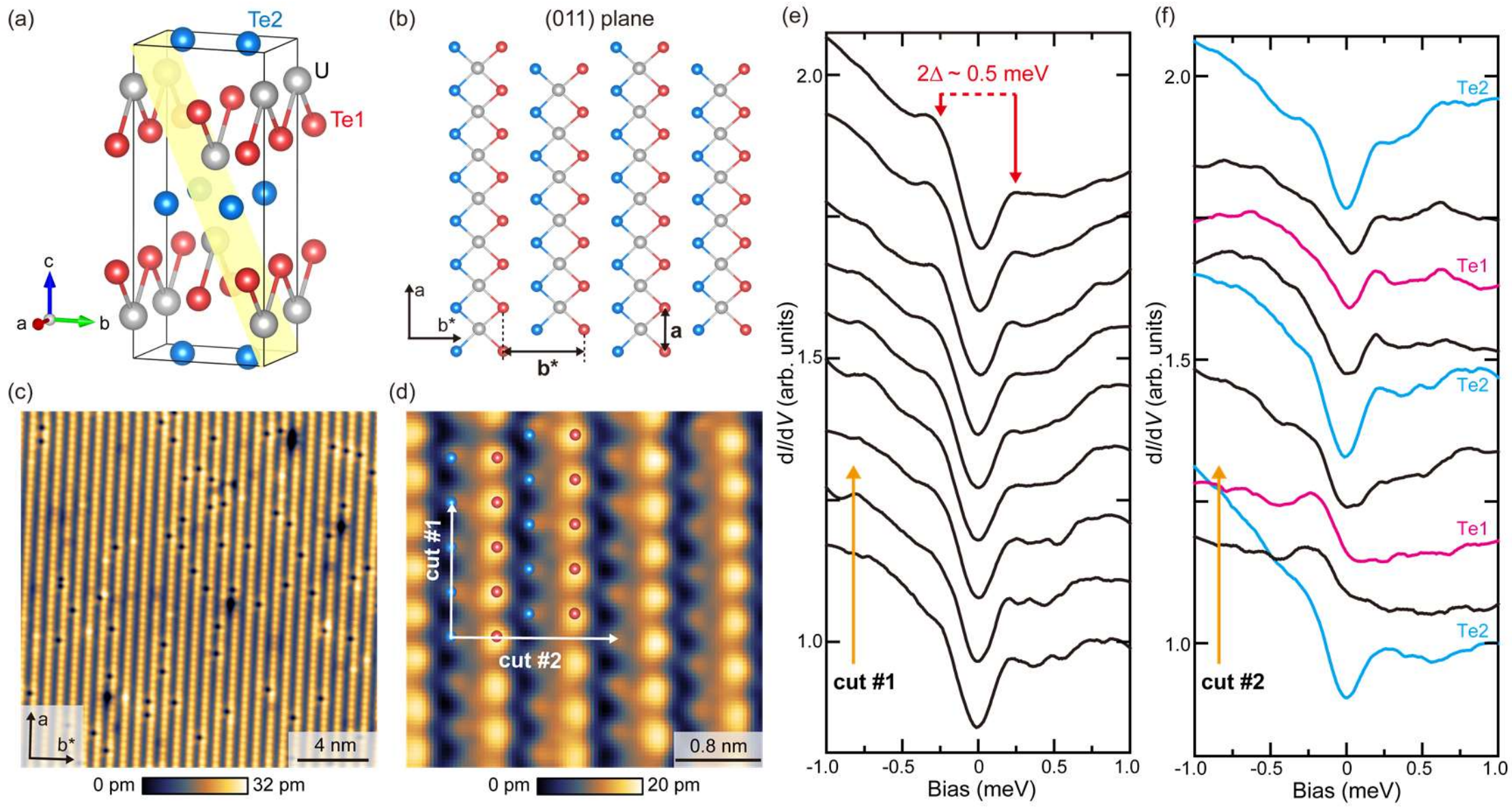

FIG. 1. Crystal structure and superconducting gap spectra of $UTe_2$. (a) Crystal structure of $UTe_2$. The easy-cleave plane is (011) plane as indicated by the yellow shaded plane. (b) Projected atomic structure of (011) plane. (c) Typical topographic image of exposed $UTe_2$ (011) surface. (d) Zoomed-in view of atomic lattice with higher resolution, with the schematic Te1 and Te2 atoms overlaid on top. (e),(f) d$I$/d$V$ spectra taken along trajectories of cuts #1 and #2 as indicated in panel (d), which are vertically shifted for clarity. Measurement conditions: (c) $V_b$ = 200 mV, $I_t$ = 50 pA; (d) $V_b$ = 100 mV, $I_t$ = 100 pA; (e,f) $V_b$ = 1 mV, $I_t$ = 100 pA, $\Delta V$ = 50 μV.

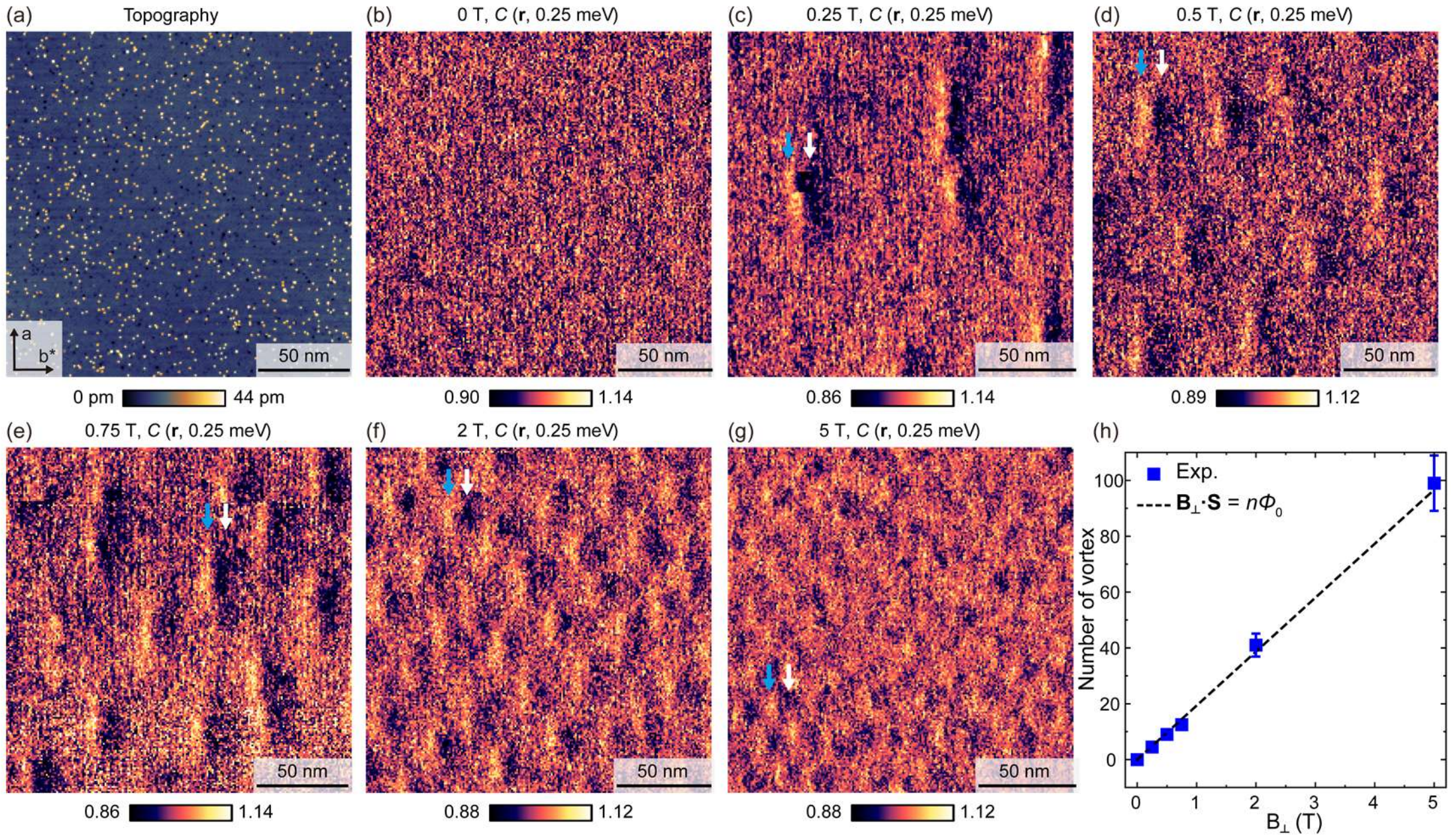

FIG. 2. Vortex maps on $UTe_2$ (011) surface under various $\mathbf{B}_\perp$. (a) Topographic image of the selected sample region for vortex mapping. (b)-(g) Normalized vortex maps under various $\mathbf{B}_\perp$. (h) Evolution of vortex number as a function of $B_\perp$, with theoretical values shown by the dashed line. Measurement conditions: (a) $V_b$ = 100 mV, $I_t$ = 30 pA; (b,d,f,g) $V_b$ = -1.5 mV, $I_t$ = 80 pA, $\Delta V$ = 100 μV; (c,e) $V_b$ = -1.5 mV, $I_t$ = 400 pA, $\Delta V$ = 80 μV.

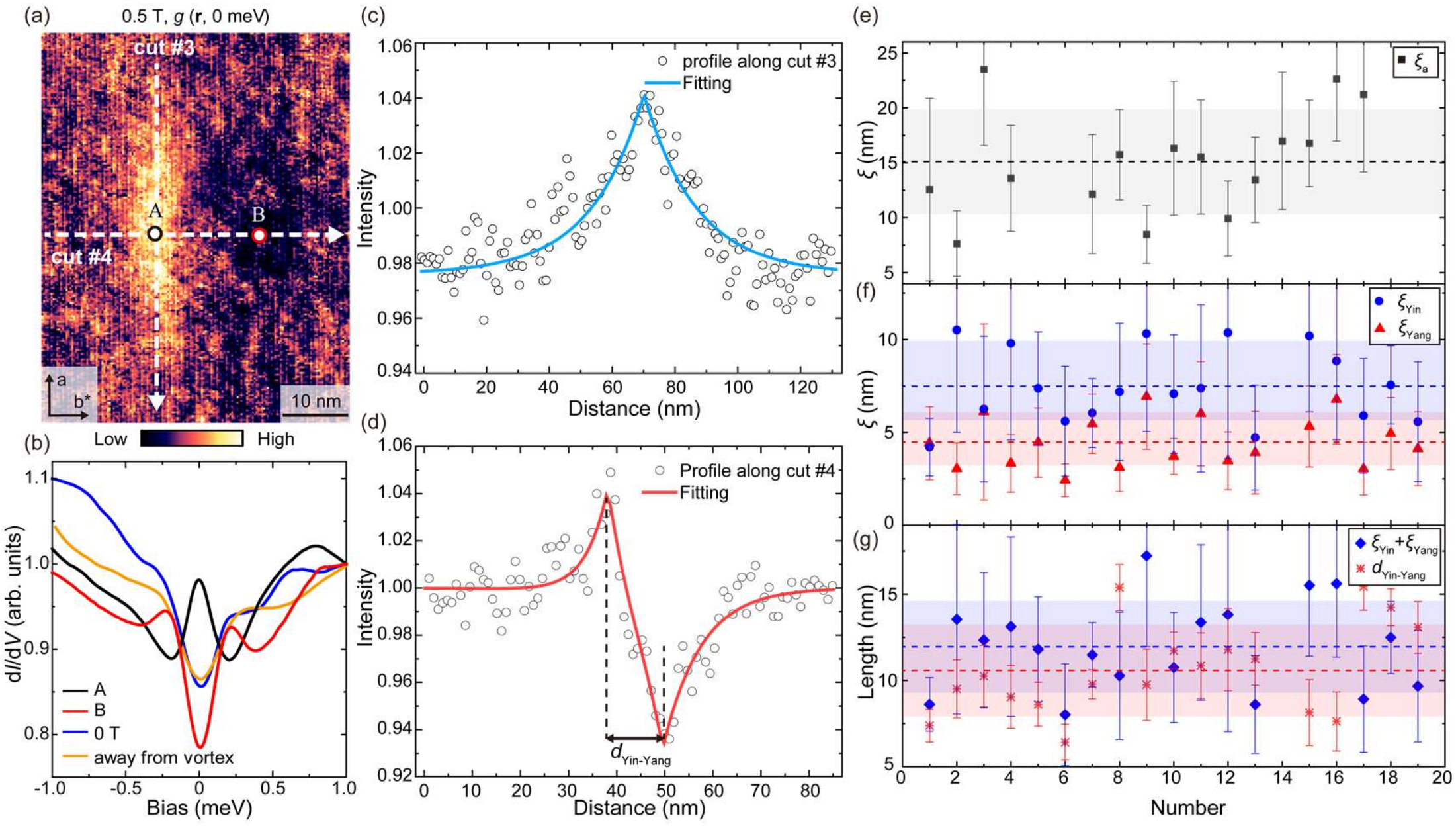

FIG. 3. Detailed properties of Yin-Yang vortex. (a) Detailed distribution of a single Yin-Yang vortex. (b) Typical d$I$/d$V$ spectra collected at the Yin and Yang parts of the vortex, as marked out by the red and black dots in panel (a). The spectra measured at the locations away from the vortex and under zero magnetic field are listed as well for comparison. (c),(d) Exponential fits to the line profiles of the vortex, taken along the trajectories of cuts #3 and #4 in panel (a). (e)-(g) Statistics of the fitted Ginzburg-Landau coherence lengths along the two lattice directions and $d_{\text{Yin-Yang}}$ for vortices measured under $B_{\perp}$ = 0.25-0.75 T. The value of the coherence length is fitted by the least square method, with the error bar at each point corresponding to the 95% confidence interval. The dashed lines indicate the averaged values of these parameters, and the shadows represent twice the standard deviation of the fitted values. Measurement conditions: (a) $V_b$ = -1.5 mV, $I_t$ = 400 pA, $\Delta V$ = 80 μV; (b) $V_b$ = 1 mV, $I_t$ = 100-400 pA, $\Delta V$ = 50 μV.

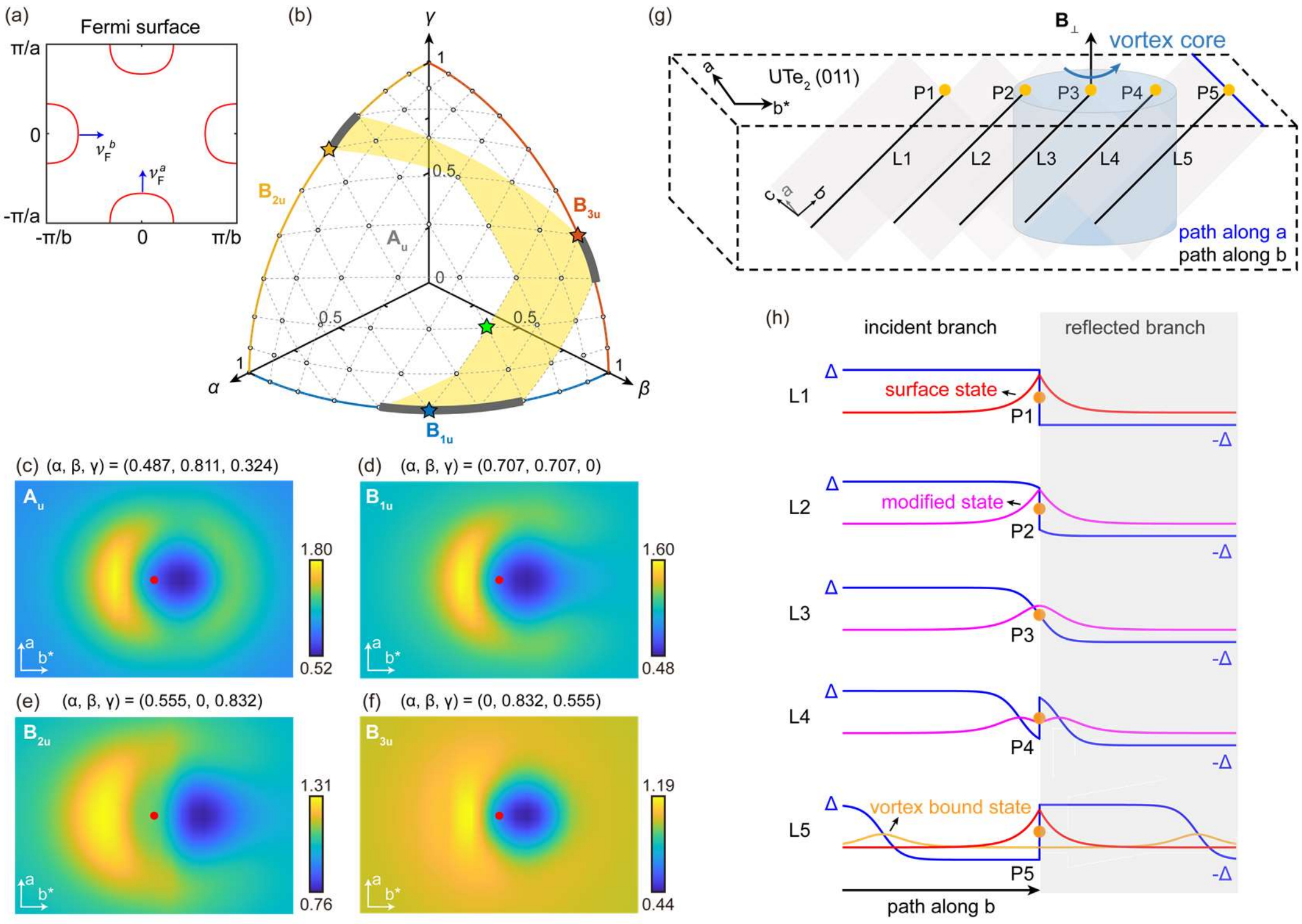

FIG. 4. Numerical results of zero-energy LDOS on $UTe_2$ (011) surface for the four types of spin-triplet pairing symmetries. (a) Cylindrical Fermi surface used for numerical simulations, with the Fermi velocity along ***a***- and ***b***-axes marked out by the blue arrows. (b) Phase diagram for the appearance of Yin-Yang vortex patterns as functions of the pairing parameters $(\alpha, \beta, \gamma)$. For simplicity, only 1/8 sphere in the first quadrant is drawn, while the cases in the other 7/8 sphere are symmetric. The blue, yellow, and orange curves represent the normalized pairing parameter spaces for $B_{1u}$, $B_{2u}$, and $B_{3u}$ pairing in general, or for $A_u$ pairing with gap nodes in accidental, while the sphere surface they enclose represents the normalized parameter space of $A_u$ pairing with a full gap. The open circles on the sphere surface denote the sampled points for our simulations. The yellow shaded strip on the sphere surface highlights the parameter regime for fully gapped $A_u$ pairing that can produce Yin-Yang vortex pattern, and its boundaries in thick gray represent the parameter regimes that can produce Yin-Yang vortex pattern for $B_{1u}$, $B_{2u}$, and $B_{3u}$ pairing with nodal gaps, or for $A_u$ pairing with accidental gap nodes. Four pentagrams mark the parameter sets used for panels (c)–(f). (c)-(f) Numerical results of zero-energy LDOS on $UTe_2$ (011) surface for the four types of pairing symmetries, with the vortex center indicated by the red dot in each plot. (g) Sketch of the typical classical paths with Fermi velocity $\boldsymbol{v}_F$ along ***b***-axis (black lines L1 to L5) and along ***a***-axis (blue line). The endpoints of paths L1-L5 on the (011) surface are labeled as P1-P5, respectively. (h) Sketch of pairing potential Δ (blue curve) along paths L1-L5 by mapping each path to a full chain, with the gray shaded region indicating the reflected branch. The zero-energy surface state (red curve) and its modified state (magenta curve), as well as the vortex bound state (orange curve), are schematically plotted as well for each path.